\documentclass[reprint,pdflatex,sn-nature]{sn-jnl}% Style for submissions to Nature Portfolio journals 

\usepackage{graphicx}%
\usepackage{multirow}%
\usepackage{subfigure}

\usepackage{lmodern}

\usepackage{amsmath,amssymb,amsfonts}%
\usepackage{wasysym}
\usepackage{amsthm}%
\usepackage{mathrsfs}%
\usepackage[title]{appendix}%
\usepackage{xcolor}%
\usepackage{textcomp}%
\usepackage{manyfoot}%
\usepackage{booktabs}%
\usepackage{algorithm}%
\usepackage{algorithmicx}%
\usepackage{algpseudocode}%
\usepackage{listings}%
\usepackage{placeins}

\usepackage{bm}

\usepackage[version=4]{mhchem}

\usepackage{siunitx} 
\usepackage[utf8]{inputenc}
\usepackage[T1]{fontenc}

\usepackage{xurl}

\renewcommand{\subref}[2]{%
	\hyperref[#1]{\ref{#1}#2}%
}

\theoremstyle{thmstyleone}%
\theoremstyle{thmstyletwo}%

\theoremstyle{thmstylethree}%

\unnumbered% uncomment this for unnumbered level heads

\newcounter{extendedfigure}

\newcommand{\BeginExtendedfigures}{
  \setcounter{extendedfigure}{0}
  \renewcommand{\thefigure}{\arabic{extendedfigure}}
  \let\origfigure\figure
  \let\endorigfigure\endfigure
  \renewenvironment{figure}
    {\refstepcounter{extendedfigure}\origfigure}
    {\endorigfigure}
}

\newcommand{\EndExtendedfigures}{
  \renewcommand{\thefigure}{\arabic{figure}}
  \let\figure\origfigure
  \let\endfigure\endorigfigure
}

\begin{document}

\title[Picosecond localization dynamics following ultrafast nanoscale magnetic switching]{Picosecond localization dynamics following ultrafast nanoscale magnetic switching}


\author[1,2,3]{\fnm{Daniel} \sur{Metternich}}

\author[3]{\fnm{Michael} \sur{Schneider}}

\author[4]{\fnm{Giuseppe} \sur{Mercurio}}

\author[5]{\fnm{Torstein} \sur{Hegstad}}

\author[6]{\fnm{Marcel} \sur{Möller}}

\author[1,2]{\fnm{Riccardo} \sur{Battistelli}}

\author[3]{\fnm{Christopher} \sur{Klose}}

\author[1]{\fnm{Steffen} \sur{Wittrock}}

\author[1,2]{\fnm{Manas R.} \sur{Patra}}

\author[1,2]{\fnm{Krishnanjana} \sur{Puzhekadavil Joy}}

\author[1,3]{\fnm{Victor} \sur{Deinhart}}

\author[1]{\fnm{Sascha} \sur{Petz}}

\author[1]{\fnm{Karel} \sur{Prokes}}

\author[1]{\fnm{Sebastian} \sur{Wintz}}

\author[1]{\fnm{Markus} \sur{Weigand}}

\author[3]{\fnm{Wolfgang-Dietrich} \sur{Engel}}

\author[3]{\fnm{Themistoklis} \sur{Sidiropoulos}}

\author[3]{\fnm{Ingo} \sur{Will}}

\author[3,7]{\fnm{Stefan} \sur{Eisebitt}}

\author[4]{\fnm{Robert E.} \sur{Carley}}

\author[4]{\fnm{Laurent} \sur{Mercadier}}

\author[4]{\fnm{Justine} \sur{Schlappa}}

\author[4]{\fnm{Martin} \sur{Teichmann}}

\author[4]{\fnm{Andreas} \sur{Scherz}}

\author[6]{\fnm{Sergey} \sur{Zayko}}

\author[8,2]{\fnm{Boris V.} \sur{Sorokin}}

\author[9]{\fnm{Kai} \sur{Bagschik}}

\author[6]{\fnm{Claus} \sur{Ropers}}

\author[5]{\fnm{Johan H.} \sur{Mentink}}

\author[3]{\fnm{Bastian} \sur{Pfau}}

\author*[1,2]{\fnm{Felix} \sur{Büttner}}\email{felix.buettner@helmholtz-berlin.de}

\affil[1]{\orgname{Helmholtz-Zentrum Berlin für Materialien und Energie GmbH}, \orgaddress{\city{Berlin}, \postcode{14109}, \country{Germany}}}

\affil[2]{\orgdiv{Experimental Physics V, Center for Electronic Correlations and Magnetism}, \orgname{University of Augsburg}, \orgaddress{\city{Augsburg}, \postcode{86159}, \country{Germany}}}

\affil[3]{\orgname{Max Born Institute for Nonlinear Optics and Short Pulse Spectroscopy}, \orgaddress{\city{Berlin}, \postcode{12489}, \country{Germany}}}

\affil[4]{\orgname{European XFEL}, \orgaddress{\city{Schenefeld}, \postcode{22869}, \country{Germany}}}

\affil[5]{\orgname{Radboud Universiteit}, \orgaddress{\city{Nijmegen}, \postcode{6525 XZ}, \country{Netherlands}}}

\affil[6]{\orgname{Max-Planck-Institut für Multidisziplinäre Naturwissenschaften}, \orgaddress{\city{Göttingen}, \postcode{37077}, \country{Germany}}}

\affil[7]{\orgname{Technische Universität Berlin, Institut für Physik und Astronomie}, \orgaddress{\city{Berlin}, \postcode{10623}, \country{Germany}}}

\affil[8]{\orgname{Paul Scherrer Institut}, \orgaddress{\city{Villigen}, \postcode{5232}, \country{Switzerland}}}

\affil[9]{\orgname{Deutsches Elektronen-Synchrotron DESY}, \orgaddress{\city{Hamburg}, \postcode{22607}, \country{Germany}}}

\abstract{
Ultrashort laser pulses provide the fastest known way to switch magnetic order. Such excitation commonly creates nanometer-scale domains, even after homogeneous illumination when the position of nucleated domains is not externally defined. However, the physics of domain localization during such ultrafast phase transitions remains unresolved. Here, we use shot-resolved pump–probe resonant x-ray scattering together with a material featuring a periodically modulated magnetic anisotropy landscape to track, in real time, the laser-driven nucleation and localization of nanometer-scale spin textures. We find that nucleation and localization are two distinct processes. Nucleation occurs homogeneously via fluctuations at early times, whereas spatially periodic structures emerge only later and, under suitable conditions, localize in less than one nanosecond. Real-space simulations show that this localization is governed by strong lateral variations in spin-texture lifetimes. Our results demonstrate that ultrafast phase-transition dynamics fundamentally differ from conventional transitions, yet still can be controlled through moderate nanometer-scale tailoring of the energy landscape.
}


\maketitle

Fundamental interactions in physics can be probed at high energies or---particularly in solid-state systems---at ultrashort timescales. Ultrafast demagnetization, for example, has revealed key energy dissipation channels~\cite{beaurepaireUltrafastSpinDynamics1996}; all-optical magnetic switching has provided insight into the complex exchange interactions between ferrimagnetic sublattices~\cite{stanciuSubpicosecondMagnetizationReversal2007,raduTransientFerromagneticlikeState2011}; and time-resolved studies of metal–insulator transitions have elucidated the interplay between electronic and lattice dynamics~\cite{wallUltrafastChangesLattice2012}. Beyond couplings between abstract degrees of freedom---such as spin, lattice, and electronic temperatures---ultrafast dynamics are also ubiquitously accompanied by nanometer-scale spatial heterogeneity~\cite{gravesNanoscaleSpinReversal2013,johnsonUltrafastXrayImaging2023,metternichDefectsMagneticDomain2025a}, the physics of which remains poorly understood. A particularly fundamental open question concerns domain localization during nanoscale ultrafast phase transitions: what are the driving forces, equations of motion, and characteristic timescales that determine the eventual position of nanoscale domains of the new phase following excitation? The implications are profound---both fundamentally, since spatial heterogeneity introduces new gradient and nonlocal terms into the Hamiltonian, and technologically, as miniaturization depends critically on spatial control.

Laser-induced switching of ferromagnetic materials exemplifies an ultrafast phase transition that often results in nanoscale domains below the size of the incident beam and even its wavelength. These domains and their spatial arrangement play a decisive role in widely studied helicity-dependent all-optical switching (HD-AOS)~\cite{khusyainovLaserinducedHelicityTexturedependent2024}. Yet, the dynamics that defines the position of domains without extrinsic guiding remains unclear. In certain perpendicular magnetic ferromagnetic multilayers, the laser-produced domains are magnetic skyrmions~\cite{jeCreationMagneticSkyrmion2018,buttnerObservationFluctuationmediatedPicosecond2021}---spin textures of well-defined topology and intrinsically-stabilized shape, which constitute the most basic domain states in out-of-plane magnetic systems. Early systematic studies reported skyrmions to form at random positions after ultrafast laser excitation~\cite{buttnerObservationFluctuationmediatedPicosecond2021,gerlingerApplicationConceptsUltrafast2021}, suggesting that ultrafast magnetic switching is governed solely by uniform magnetic interactions---primarily exchange---without significant influence from local variations in the micromagnetic energy landscape~\cite{buttnerObservationFluctuationmediatedPicosecond2021}. Similar observations were made for nanometer-scale topological textures in in-plane magnetic films~\cite{eggebrechtLightInducedMetastableMagnetic2017}. More recent experiments, however, have challenged this interpretation by demonstrating deterministic spatial control of skyrmion formation after laser excitation through ion-beam patterning of the underlying magnetic anisotropy~\cite{kernDeterministicGenerationGuided2022,kernControlledFormationSkyrmion2025}. This development provides a unique opportunity to clarify the mechanisms and timescales governing localization during ultrafast phase transitions, and to determine when and how spatial constraints and nonlocal interactions shape the emergent nanoscale order.

\begin{figure}[p!]
	\centering
	\includegraphics[width=250pt]{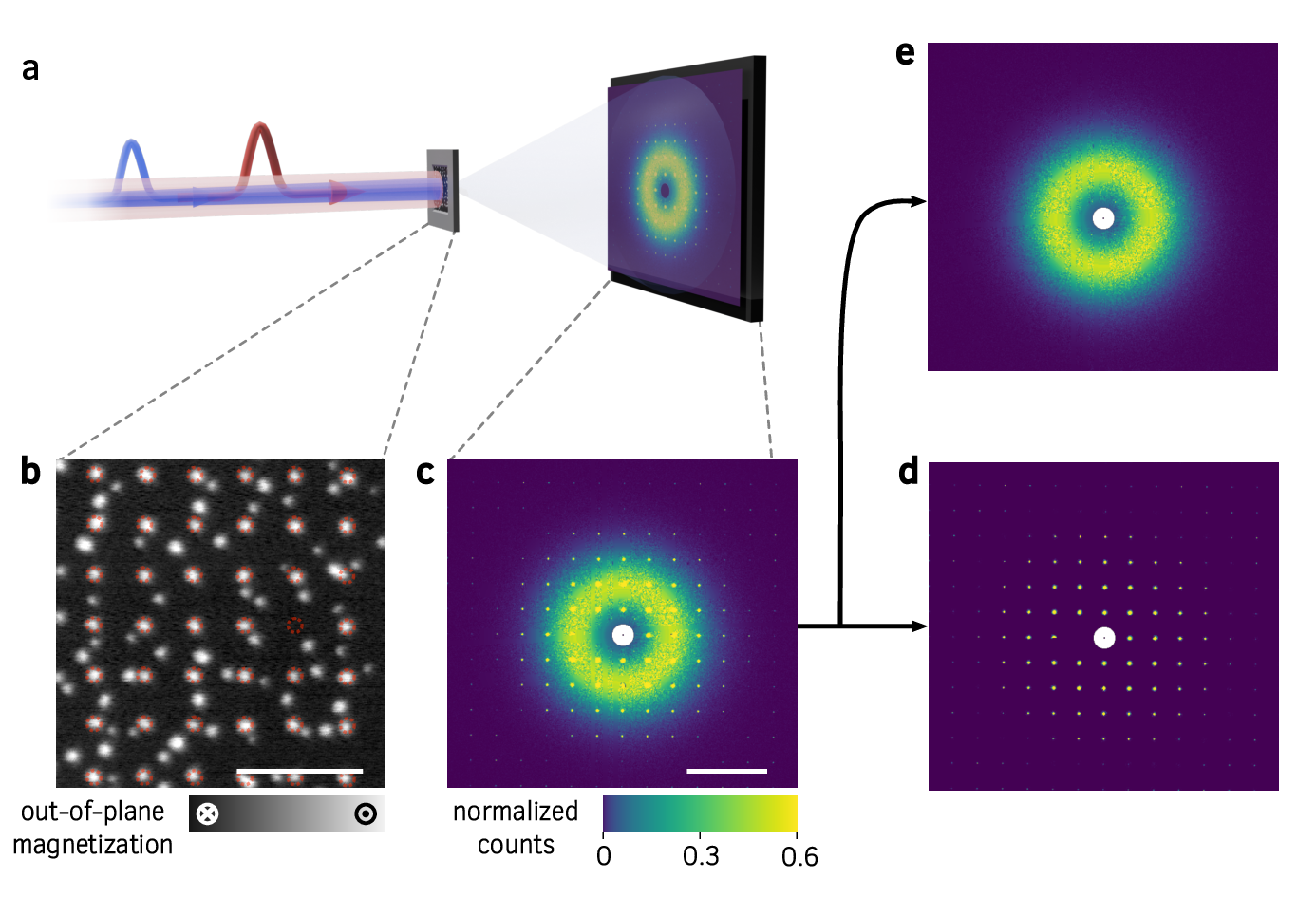}
	\caption{ \textbf{Concept of measuring localization in real time.} \textbf{a} Schematic of the time-resolved SAXS experiment. Red: optical femtosecond pump beam, blue: x-ray probe beam, gray: sample, black: camera detector. \textbf{b} Real-space configuration of nanometer-scale magnetic domains, partially localized on a grid of artificial pinning sites with locally reduced perpendicular anisotropy (indicated in red). Recorded with STXM, scalebar \qty{1}{\micro\meter}. \textbf{c} Measured scattering pattern from an irradiated sample, showing both an isotropic ring and square-symmetric Bragg peaks. The ring originates from disordered magnetic textures, whereas the ordered skyrmions aligned to the ion-beam pattern give rise to Bragg peaks at the reciprocal lattice points of the grid. Scalebar \qty{0.05}{\per\nano\meter}. \textbf{d} Isolated Bragg-peak contribution obtained by separating the skyrmion-lattice signal from the isotropic background, highlighting only the reciprocal-space signature of the ordered skyrmions. \textbf{e} Isotropic scattering contribution from disordered magnetic features, including unpinned skyrmions.}
	\label{fig:1}
\end{figure}

Here, we use resonant small-angle x-ray scattering (SAXS) at an x-ray free-electron laser (XFEL) to track the spatiotemporal evolution of magnetic textures following femtosecond laser excitation. The key innovation of our study lies in employing a magnetic multilayer with a periodic, ion-beam–induced modulation of the magnetic anisotropy. This approach allows us to quantitatively distinguish between homogeneously distributed and spatially localized magnetic textures in the measured Fourier-space data (Fig.~\ref{fig:1}). Our measurements reveal that nucleation and localization occur on distinct timescales: nucleation dominates the first few hundred picoseconds after excitation, while localization emerges only subsequently. By combining atomistic spin dynamics simulations with micromagnetic modeling, we disentangle high-temperature, fluctuation-driven nucleation from micromagnetic stabilization processes that govern localization. Together, these results establish an experimentally validated multiscale framework for describing nanoscale dimensions of ultrafast phase transitions, and open the door to exploiting such processes for controlled nanoscale device functionality.

\section{Engineering of the material} 

The base material of our study is a [Co/Pt]$_{\times 15}$ ferromagnetic multilayer, which exhibits robust and well-characterized nucleation of nanometer-scale magnetic textures (especially skyrmions) following ultrafast laser excitation above a critical fluence~\cite{buttnerObservationFluctuationmediatedPicosecond2021,gerlingerApplicationConceptsUltrafast2021,gerlingerRobustScenarioGeneration2022}. In similar multilayers, controlled localization of magnetic textures at ion-irradiated regions has been achieved through local reductions in interfacial anisotropy and, where present, the Dzyaloshinskii–Moriya interaction (DMI)~\cite{sapozhnikovArtificialDenseLattice2016,jugeHeliumIonsPut2021,kernDeterministicGenerationGuided2022,dejongControllingMagneticSkyrmion2023,gusevModificationInterfacialDzyaloshinskii2021,dejongLocalControlMagnetic2022,ahrensSkyrmionsControlFIB2023,kernControlledFormationSkyrmion2025}. Building on the general concept of energy-landscape engineering, we introduced a periodic modulation of the magnetic anisotropy by patterning the multilayer with a focused Ga$^+$ ion beam into circular \qty{100}{nm}-diameter ``dots'' (see Methods). Magneto-optical Kerr and scanning transmission x-ray microscopy (STXM) measurements confirm that ion irradiation increases the out-of-plane saturation field and decreases the stripe-domain periodicity, consistent with a local reduction in perpendicular magnetic anisotropy (Extended Data Fig.~\ref{fig:xfel-material}).

\begin{figure}[p!]
	\centering
	\includegraphics[width=250pt]{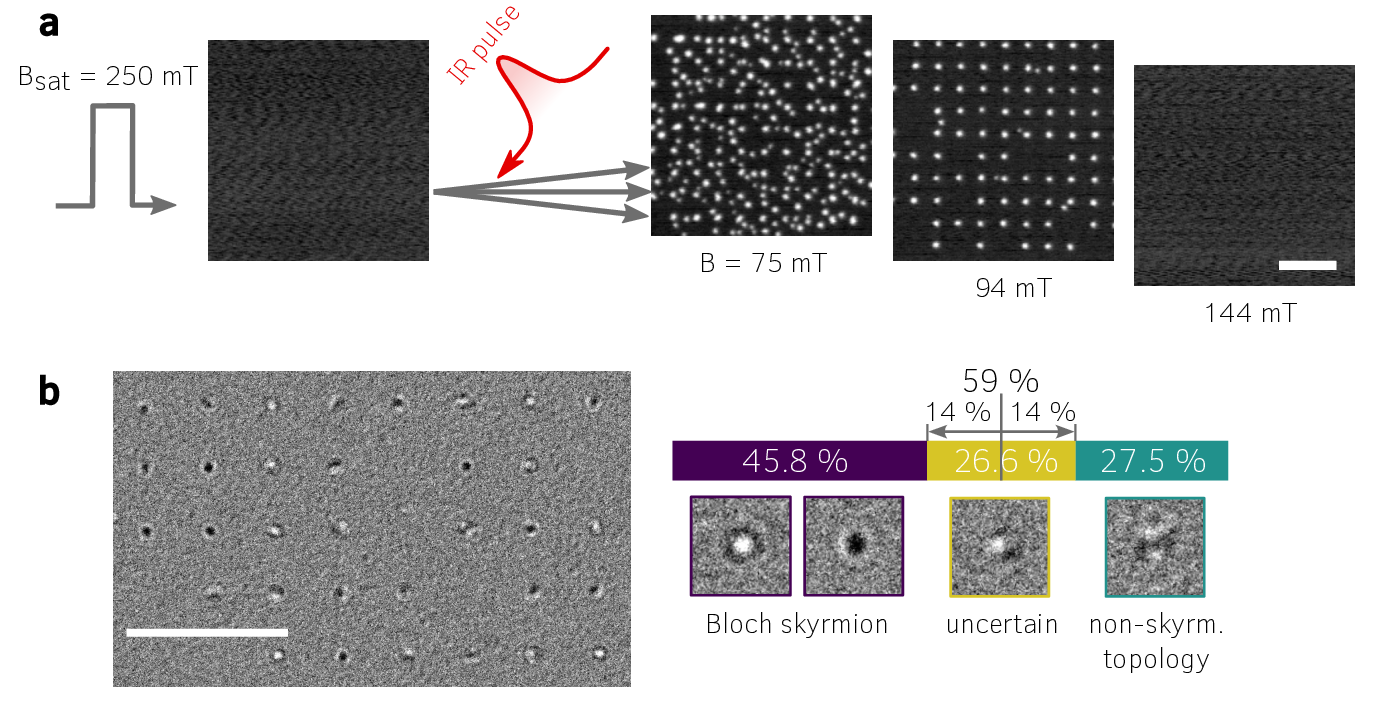}
	\caption{\textbf{Real space localization of nanometer-scale domains at ion-irradiated sites.} \textbf{a} Magnetic states after single-shot laser irradiation for various applied magnetic fields~$B$. Each state was created from saturation. Recorded with STXM. \textbf{b} Lorentz-TEM contrast of circular domains localized on the grid of ion-irradiated sites. The bar diagram depicts the percentage of textures identified as skyrmions, as non-skyrmionic bubbles, and uncertain topology, along with examples of each category. Scalebars, \qty{1}{\micro\meter}.}
	\label{fig:2}
\end{figure}

To test the localization capability, the patterned film was imaged after it was magnetically saturated, brought to an out-of-plane holding field above the spontaneous nucleation threshold, and then exposed to a single ultrafast laser pulse. STXM images (Fig.~\subref{fig:2}{a}) reveal three regimes: at low fields (\qtyrange{30}{75}{mT}), dense, disordered arrays of circular domains form; at high fields (\qty{>130}{mT}), the film remains saturated; and at intermediate fields, each irradiated dot hosts a single domain while the surrounding area remains uniformly magnetized. Lorentz transmission electron microscopy (L-TEM) (Fig.~\subref{fig:2}{b}) shows that \qty{59 +- 14}{\percent} (confidence interval) of domains in the dots are skyrmions, with the remainder being non-topological or of higher-order topologies. While this observation is contrary to earlier experiments on the non-irradiated film~\cite{buttnerObservationFluctuationmediatedPicosecond2021}, it is consistent with reports of energetically degenerate topological states in similar low-anisotropy, achiral multilayers~\cite{heiglDipolarstabilizedFirstSecondorder2021,hassanDipolarSkyrmionsAntiskyrmions2024}, and provides an opportunity to explore how topology affects the nucleation and localization dynamics. Overall, the quasi-static measurements confirm that the engineered material fulfills all requirements for the time-resolved experiments.

\section{Observation of localization dynamics in Fourier space} 

\begin{figure}[p!]
	\centering
	\includegraphics[]{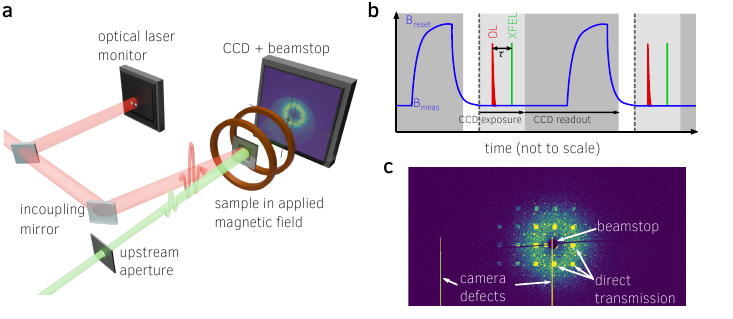}
	\caption{\textbf{Schematic of the time resolved SAXS experiment.} \textbf{a} Conceptual sketch of the experimental setup. \textbf{b} Control sequence to record time-resolved data. After resetting the magnetic state at $B_{\text{reset}}$ and consequently reducing the applied field to $B_{\text{meas}}$, the optical pump and x-ray probe pulse hit the sample with the specified time delay $\tau$. \textbf{c} Example of a raw final-state camera frame, showing magnetic scattering and the artifacts that need to be removed in post-processing.}
	\label{fig:xfel-procedure}
\end{figure}

To follow nucleation and localization dynamics, we performed time-resolved small-angle x-ray scattering (SAXS) with x-ray magnetic circular dichroism contrast at the SCS beamline of European XFEL. Before each measurement, the sample was magnetically saturated, the field reduced, and a single femtosecond laser pulse was applied followed by an XFEL probe pulse at a defined delay (Extended Data Fig.~\ref{fig:xfel-procedure}). Each single-shot scattering pattern was recorded and analyzed individually. Time-delay series were acquired for the three magnetic-field regimes identified earlier: a high field (\qty{165}{mT}) yielding a saturated final state, an intermediate field (\qty{115}{mT}) producing localization on the ion-irradiated grid, and a low field (\qty{65}{mT}) generating textures throughout the film. Each scattering pattern $I(\mathbf{q},t)$ was background-corrected, normalized, averaged over identical conditions, and decomposed into isotropic ($I_{\text{iso}}(\mathbf{q},t)$) and Bragg ($I_{\text{Bragg}}(\mathbf{q},t)$) components corresponding to non-localized and localized textures as shown in Fig.~\subref{fig:1}{c–e}. See Methods for details.

\begin{figure}[p!]
	\centering
	\includegraphics[width=\textwidth]{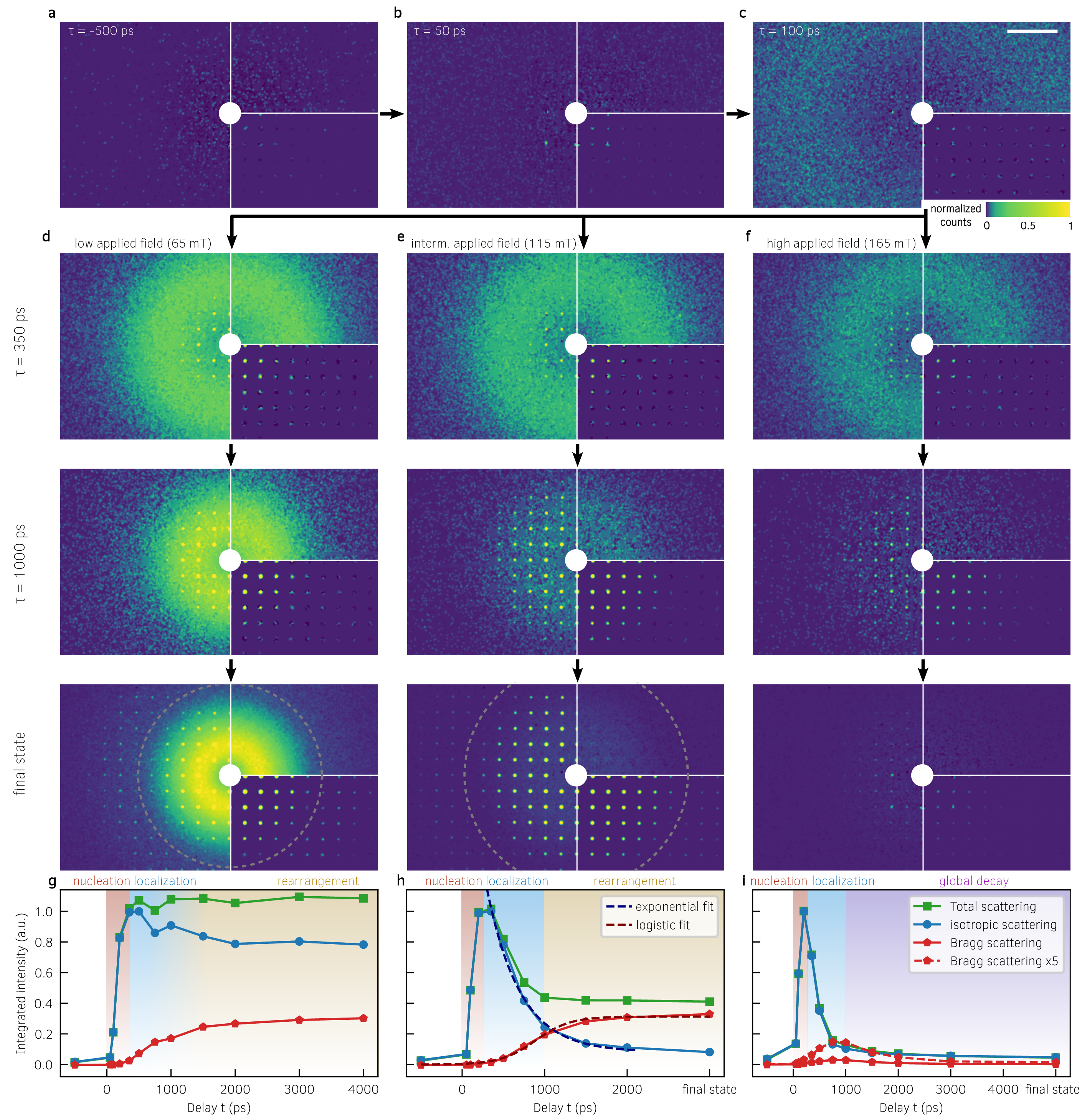}
	\caption[Evolution of the scattering signal after optical excitation.]{\textbf{Evolution of the scattering signal after optical excitation.} \textbf{a-f} 
    Scattering patterns at selected time delays after the optical pump for applied fields of 65, 115, and \qty{165}{mT}. Patterns for delays \qty{\leq100}{ps} are shown only once, as they are identical across fields. Left panels: raw background-corrected data including isotropic and Bragg scattering. The white central circle marks the beamstop region with no recorded signal. Top right: isotropic scattering. Bottom right: Bragg scattering. Dashed circles indicate the first Airy minimum of the Bragg signal. A Gaussian filter (standard deviation \qty{1.5}{px}) was applied before display for clarity. Scalebar \qty{0.05}{\per\nano\meter}. \textbf{g-i} Integrated scattering intensity, separated into isotropic and Bragg contributions, for the three applied fields. Intensities were normalized to the maximum of the isotropic signal. Background colors denote the approximate durations of the distinct periods of the dynamics. See text for details.}
	\label{fig:3}
\end{figure}

\begin{figure}[p!]
	\centering
	\includegraphics[width=\textwidth]{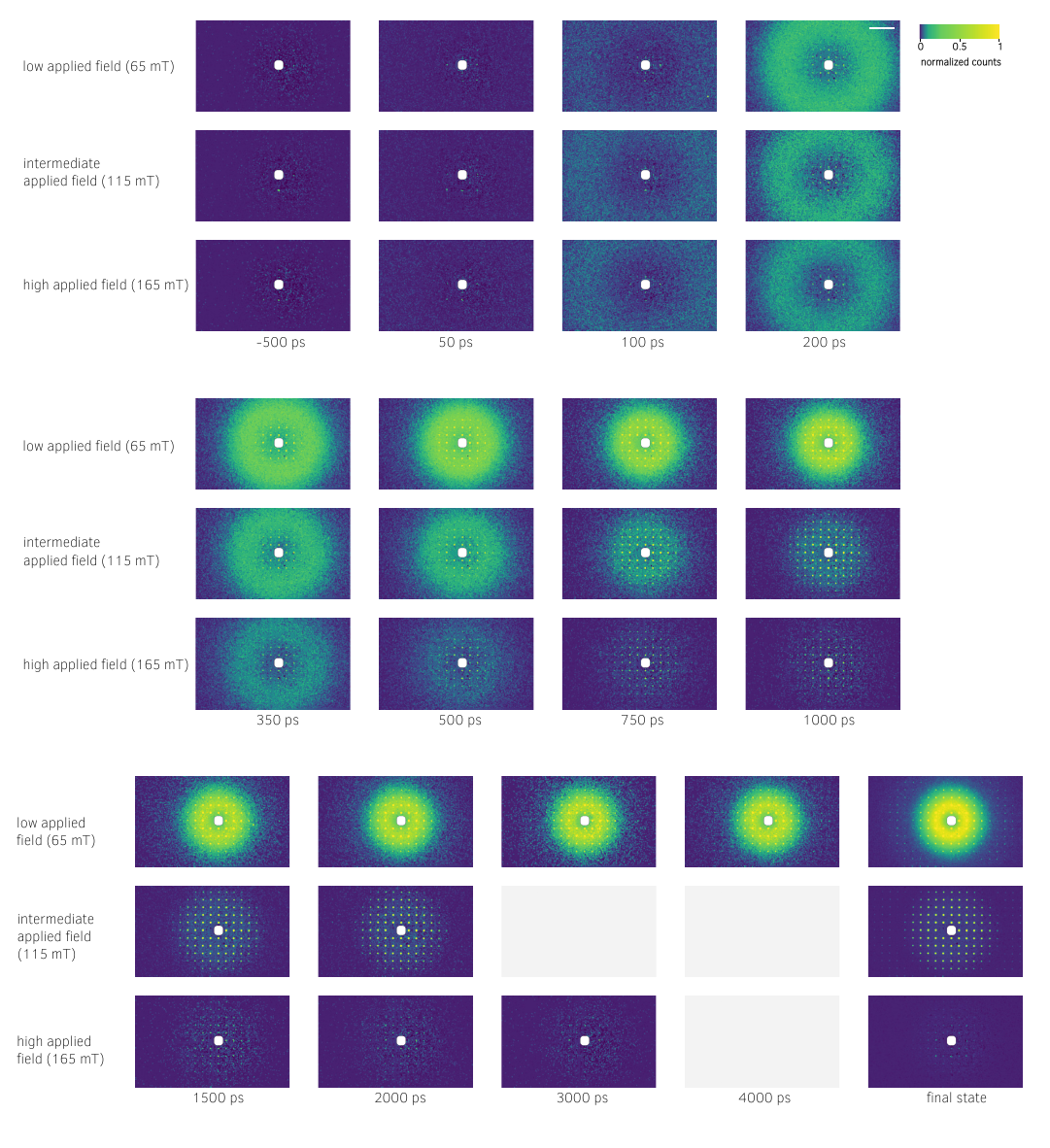}
	\caption{\textbf{Resonant magnetic scattering patterns for all recorded fields and delays.} For displaying purposes, a Gaussian filter ($\sigma=\qty{2}{px}$) was applied. Gray panels indicate that no data was recorded at that combination of delay and field. Scalebar \qty{0.05}{\per\nano\meter}.\label{fig:xfel-allsaxspatterns}}
\end{figure}

\begin{figure}[p!]
	\centering
	\includegraphics[width=\textwidth]{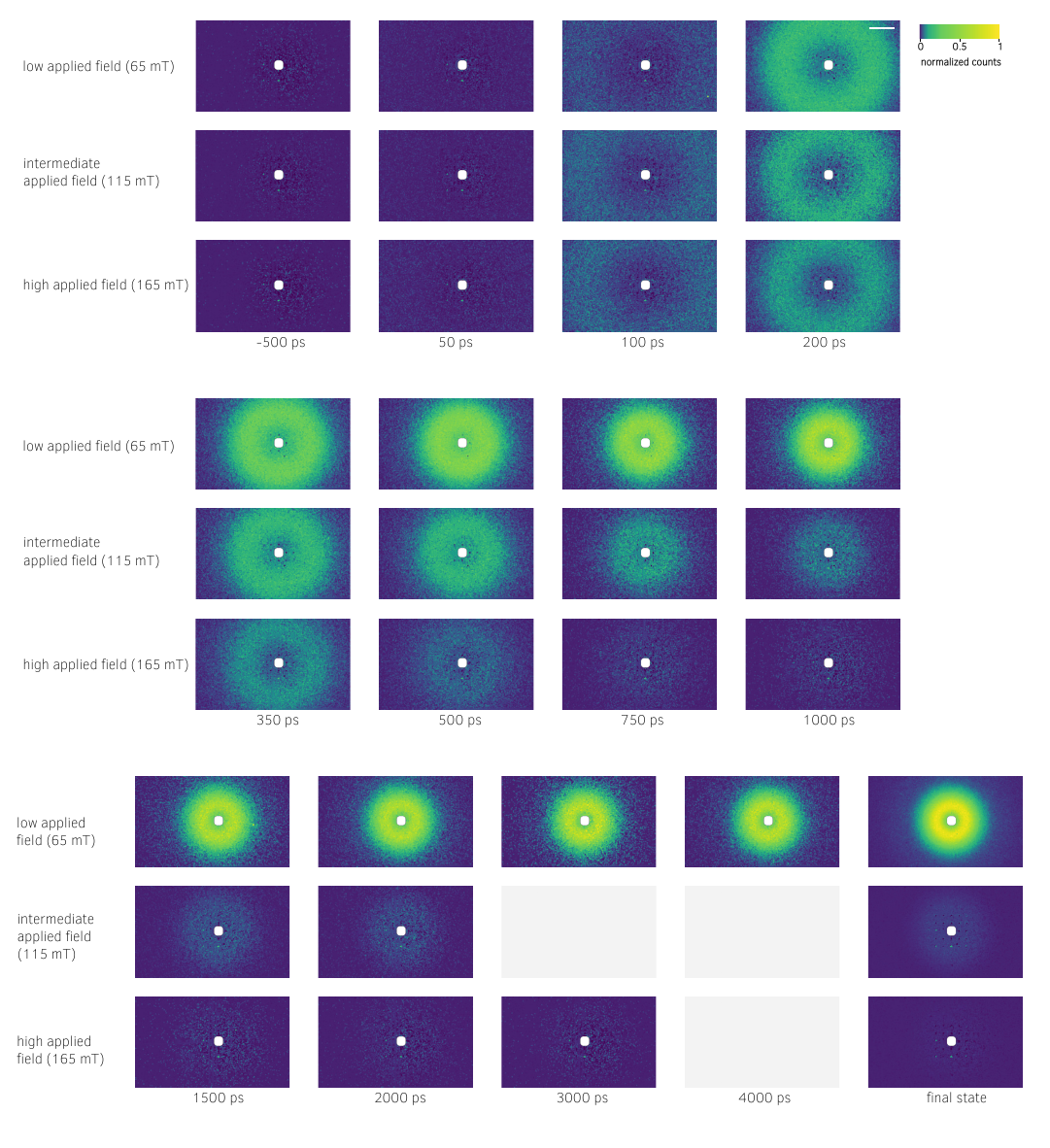}
	\caption{\textbf{Isolated isotropic magnetic scattering patterns for all recorded fields and delays.} For displaying purposes, a Gaussian filter ($\sigma=\qty{2}{px}$) was applied. Gray panels indicate that no data was recorded at that combination of delay and field. Scalebar \qty{0.05}{\per\nano\meter}.\label{fig:xfel-allsaxspatternsiso}}
\end{figure}

\begin{figure}[p!]
	\centering
	\includegraphics[width=\textwidth]{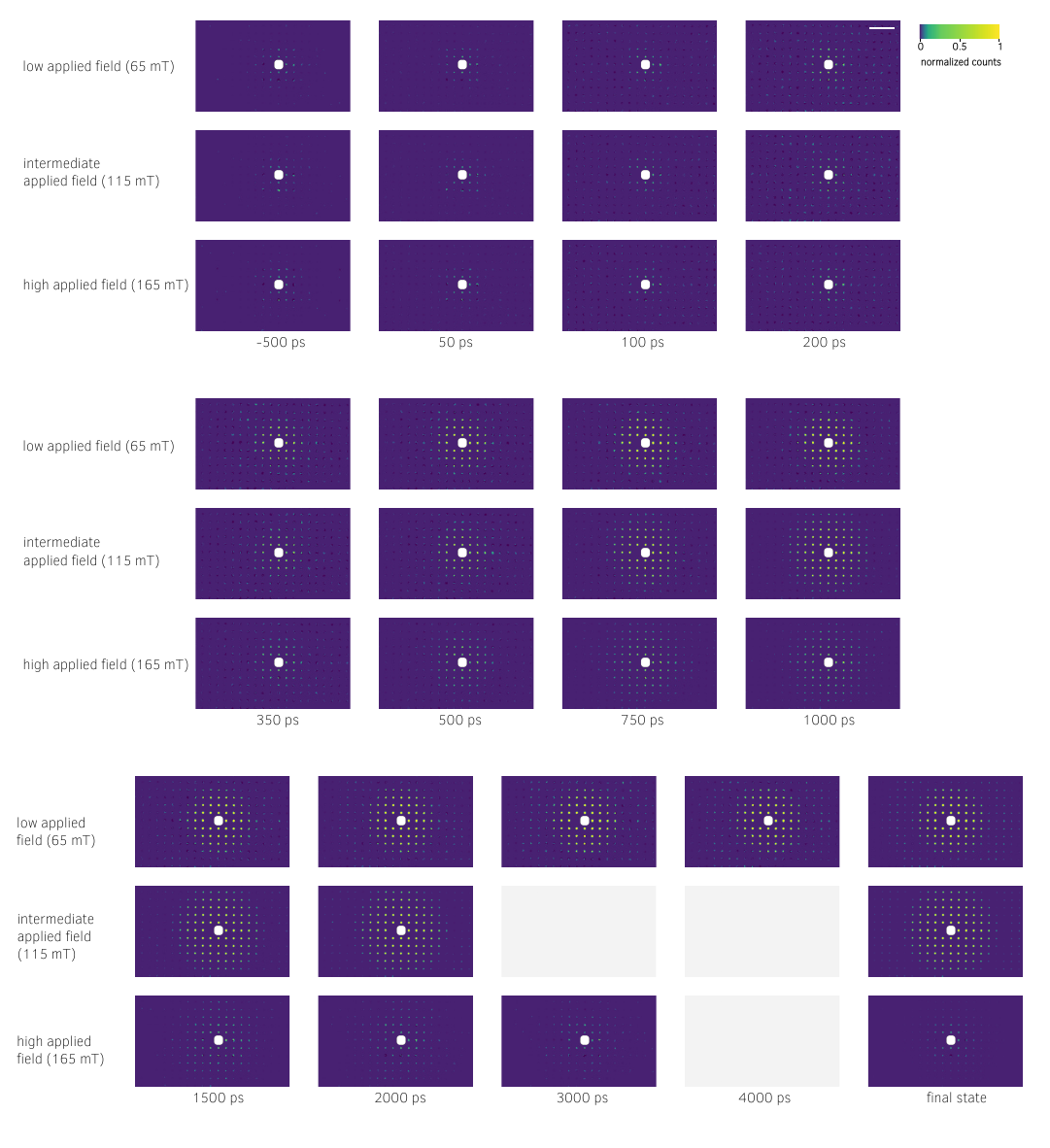}
	\caption{\textbf{Isolated Bragg scattering patterns for all recorded fields and delays.} For displaying purposes, a Gaussian filter ($\sigma=\qty{2}{px}$) was applied. Gray panels indicate that no data was recorded at that combination of delay and field. Scalebar \qty{0.05}{\per\nano\meter}.\label{fig:xfel-allsaxspatternsbragg}}
\end{figure}

The temporal evolution of the scattering signal is shown in Fig. \ref{fig:3} and Extended Data Figs.~\ref{fig:xfel-allsaxspatterns},\ref{fig:xfel-allsaxspatternsiso},\ref{fig:xfel-allsaxspatternsbragg}. Shortly after excitation, weak, uniform scattering emerges and evolves into a diffuse ring that contracts in Fourier space (Figs.~\subref{fig:3}{a–c}). This behavior, identical across all field strengths and consistent with earlier work~\cite{buttnerObservationFluctuationmediatedPicosecond2021}, reflects entry into a fluctuation regime with structures subsequently condensing into magnetic textures with defined size and spacing~\cite{buttnerObservationFluctuationmediatedPicosecond2021,suturinShortNanometerRange2023}. Bragg scattering remains negligible during this stage and up to \qty{300}{ps}, when nucleation completes (Figs.~\subref{fig:3}{d–f}) \cite{buttnerObservationFluctuationmediatedPicosecond2021}. At later delays, pronounced Bragg peaks emerge across all field regimes, with up to six visible diffraction orders, indicating long-range periodic ordering of localized textures. In the low- and intermediate-field regimes, characteristic Airy-disk minima in the Bragg intensity, arising from the texture form factor, reveal the circular symmetry of the localized textures, with field-dependent diameters of approximately \qty{89}{nm} and \qty{71}{nm}, respectively, that remain constant beyond \qty{750}{ps} (Extended Data Fig.~\ref{fig:xfel-airy}). In these regimes, the scattering patterns stabilize after roughly \qty{1}{ns} (Extended Data Figs.~\ref{fig:xfel-allsaxspatterns},\ref{fig:xfel-allsaxspatternsiso},\ref{fig:xfel-allsaxspatternsbragg}), closely matching the quasi-static real-space images (Fig.~\ref{fig:2}): pure Bragg scattering at intermediate fields and mixed isotropic–Bragg contributions at low fields. At high fields, the transiently localized textures decay slightly later, after approximately \qty{1.5}{ns}, also yielding a scattering pattern consistent with the expected quasi-static high-field state (Extended Data Figs.~\ref{fig:xfel-allsaxspatterns},\ref{fig:xfel-allsaxspatternsiso},\ref{fig:xfel-allsaxspatternsbragg}).

Localization dynamics were quantified by integrating scattering intensities over all scattering vectors $\mathbf{q}$, i.\,e. over all probed spatial frequencies. The total intensity $I(t)$ reflects the overall density of textures, while $I_{\text{iso}}(t)$ and $I_{\text{Bragg}}(t)$ represent the non-localized and localized fractions, respectively. Across all field regimes, the isotropic contribution rises rapidly after excitation and reaches a maximum at about \qty{300}{ps} (Figs.~\subref{fig:3}{g–i}). Its temporal evolution closely matches previously reported fluctuation-mediated nucleation dynamics—including the same characteristic timescale~\cite{buttnerObservationFluctuationmediatedPicosecond2021}—identifying the interval up to this maximum as the \textit{nucleation period}. The subsequent onset of Bragg-intensity growth marks the beginning of the \textit{localization period}. Bragg growth follows a single-time-constant logistic growth model—at least in the low- and intermediate-field regimes—indicating that the various spin textures inside the dots, regardless of topology, localize through similar dynamics.

The end of the localization period is marked by field-dependent signatures. At high fields, all textures eventually decay, and the resulting drop in Bragg intensity at roughly \qty{1}{ns} marks the transition to the \textit{global-decay period}. At intermediate fields, the total intensity saturates at about the same time ($\partial_t I \approx 0$), defining a \textit{rearrangement period} in which intensity is gradually transferred from isotropic to Bragg components, indicating refinement of spatial order without changes in texture number or size. At low fields, the post-\qty{1}{ns} behavior is similar, although the boundary between localization and rearrangement is less distinct because the total intensity is already nearly constant during localization.

Overall, these results demonstrate that localization begins only after the fluctuations have faded away, completes within a sub-nanosecond, and proceeds similarly for all spin textures, independent of topology.

\section{Modeling of localization dynamics in real space} 

\begin{figure}[p!]
	\centering
	\includegraphics[width=\textwidth]{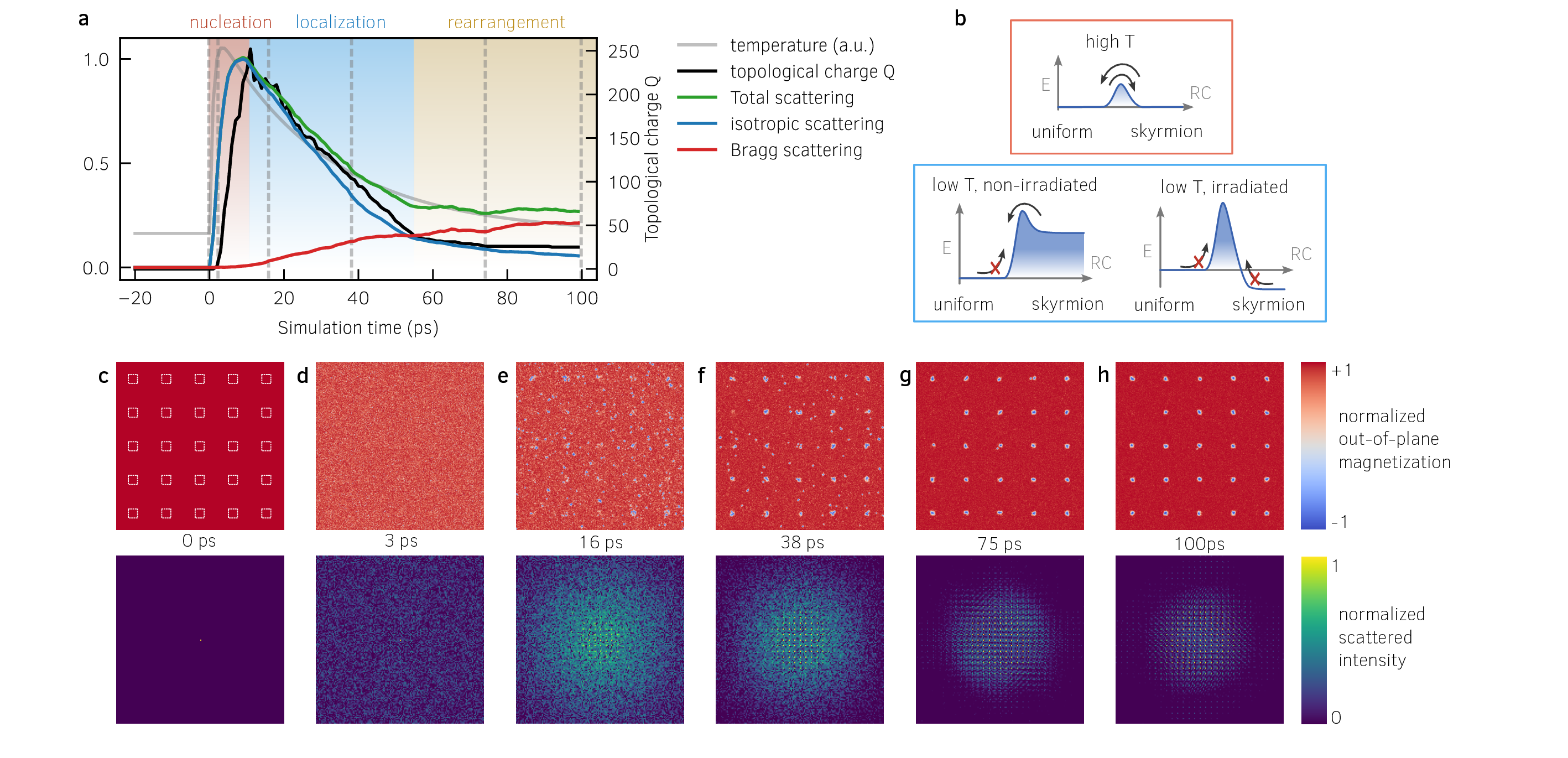}
	\caption{\textbf{Atomistic spin dynamics simulations of the nucleation process.} \textbf{a} Time evolution of total isotropic and Bragg-like scattering in the simulation, bath temperature, and total topological charge (skyrmion count). a.u., arbitrary units. Gray dashed lines indicate the time points shown in (c)–(h). Simulation times are compressed relative to the experiment. \textbf{b} Conceptual energy diagram ($E$ vs. reaction coordinate $RC$) illustrating the high-temperature nucleation regime and low-temperature decay regime inside and outside irradiated dots. \textbf{c}-\textbf{h} Snapshots of simulated real-space magnetic textures and the corresponding calculated scattering patterns. Colors indicate the out-of-plane magnetization component and scattering intensity, respectively. Dashed squares in (c) outline the regions of reduced anisotropy.}
	\label{fig:4}
\end{figure}

To uncover the real-space processes underlying the SAXS signal---and thus the mechanism of spatial localization---we performed atomistic spin-dynamics simulations. Following our previously established approach for ultrafast nucleation of topological textures~\cite{buttnerObservationFluctuationmediatedPicosecond2021}, the simulated system was reduced for computational efficiency: the film thickness was compressed to a single atomic layer, the lateral system size scaled down, and stray-field interactions were replaced by an effective Dzyaloshinskii–Moriya interaction (DMI) as the source of non-collinearity. These simplifications accelerate the simulated dynamics and thus compress the time axis relative to experiment. Optical excitation was implemented via a time-dependent thermal bath (Fig.~\subref{fig:4}{a}), and the periodic ion-irradiation pattern via a $5\times5$ array of reduced-anisotropy squares under periodic boundary conditions (Fig.~\ref{fig:4}{c}). The out-of-plane magnetic field was chosen to reproduce the experimental intermediate-field regime. Simulated out-of-plane magnetization maps were Fourier-transformed to generate SAXS patterns (in linear approximation), decomposed into isotropic and Bragg components, and integrated over $\mathbf{q}$ following the same analysis pipeline as in the experiment (see Methods).

The evolution of the $q$-integrated simulated scattering intensity (Fig.~\subref{fig:4}{a}) closely reproduces the experimental results (Fig.~\subref{fig:3}{h}). The isotropic component rises sharply and decays exponentially, while the Bragg component increases more slowly with a delayed onset; the total scattering intensity saturates at late times. The simulated 2D SAXS patterns follow the same sequence observed experimentally---uniform scattering, contraction, and subsequent emergence of sharp Bragg peaks (Figs.~\ref{fig:4}{c-h}). This close correspondence confirms that the simplified model captures the essential dynamics and enables mapping of the experimental transition times of the nucleation, localization, and rearrangement periods onto the simulation times (Fig.~\subref{fig:4}{a}).

The simulated magnetization dynamics (Figs.~\subref{fig:4}{c–h}) reveal the microscopic origin of these regimes. During the nucleation period, immediately after excitation, high-temperature fluctuations enable homogeneous formation of topological nuclei across the film (Figs.~\subref{fig:4}{d–e}) through thermal activation over a reduced topological barrier (Fig.~\subref{fig:4}{b})~\cite{buttnerObservationFluctuationmediatedPicosecond2021,liefferinkEffectiveTheoryUltrafast2025}. The total topological charge rises rapidly (Fig.~\subref{fig:4}{a}). Because the nucleation barrier depends primarily on exchange interactions~\cite{buttnerTheoryIsolatedMagnetic2018,liefferinkEffectiveTheoryUltrafast2025}, this process is largely insensitive to the anisotropy modulation. As the system cools, it enters a decay-dominated regime in which nuclei outside irradiated regions collapse, while those within persist and expand toward equilibrium size (Fig.~\subref{fig:4}{f}). This decay-mediated localization occurs during the experimentally identified localization period and arises from spatial variations in the decay barrier and associated lifetimes governed by local micromagnetic parameters. Finally, the rearrangement period is characterized by a single stable skyrmion on each dot, which gradually relax toward the centers of the irradiated spots as thermal motion decreases (Figs.~\subref{fig:4}{g–h}). Thus, the slow transfer of SAXS intensity from isotropic to Bragg components in this stage indeed reflects progressive ordering without further nucleation or annihilation.

\section{Micromagnetic understanding of localization} 

While the atomistic simulations provide an excellent qualitative description of the nucleation and localization dynamics, their computational cost prevents modeling of the full stray-field–coupled system and thus any quantitative predictions. We therefore adopt an analytical approach to test whether localization can be inferred quantitatively from the micromagnetic parameters of the material. Building on earlier indication that laser excitation produces skyrmions only when their energy is lower than that of the uniform state~\cite{gerlingerApplicationConceptsUltrafast2021}, we employ an analytical model for the energy difference between an isolated, circularly symmetric spin texture and the uniform state~\cite{buttnerTheoryIsolatedMagnetic2018}. The model provides approximate expressions for exchange, anisotropy, DMI, stray-field, and Zeeman contributions, yielding the total energy as a function of skyrmion diameter and thus the equilibrium size, energy, and barrier heights. Note that higher-order topological textures cannot be captured by the model due to the assumption of circular symmetry; however, if they are energetically degenerate to skyrmions~\cite{heiglDipolarstabilizedFirstSecondorder2021,hassanDipolarSkyrmionsAntiskyrmions2024}, the model also applies to them. Thus, the model predicts whether a nanometer-scale magnetic texture in a given local environment exhibits a global, local, or no energy minimum relative to the uniform state---corresponding to stable, metastable, or unstable configurations.

\begin{figure}[p!]
	\centering
	\includegraphics[width=250pt]{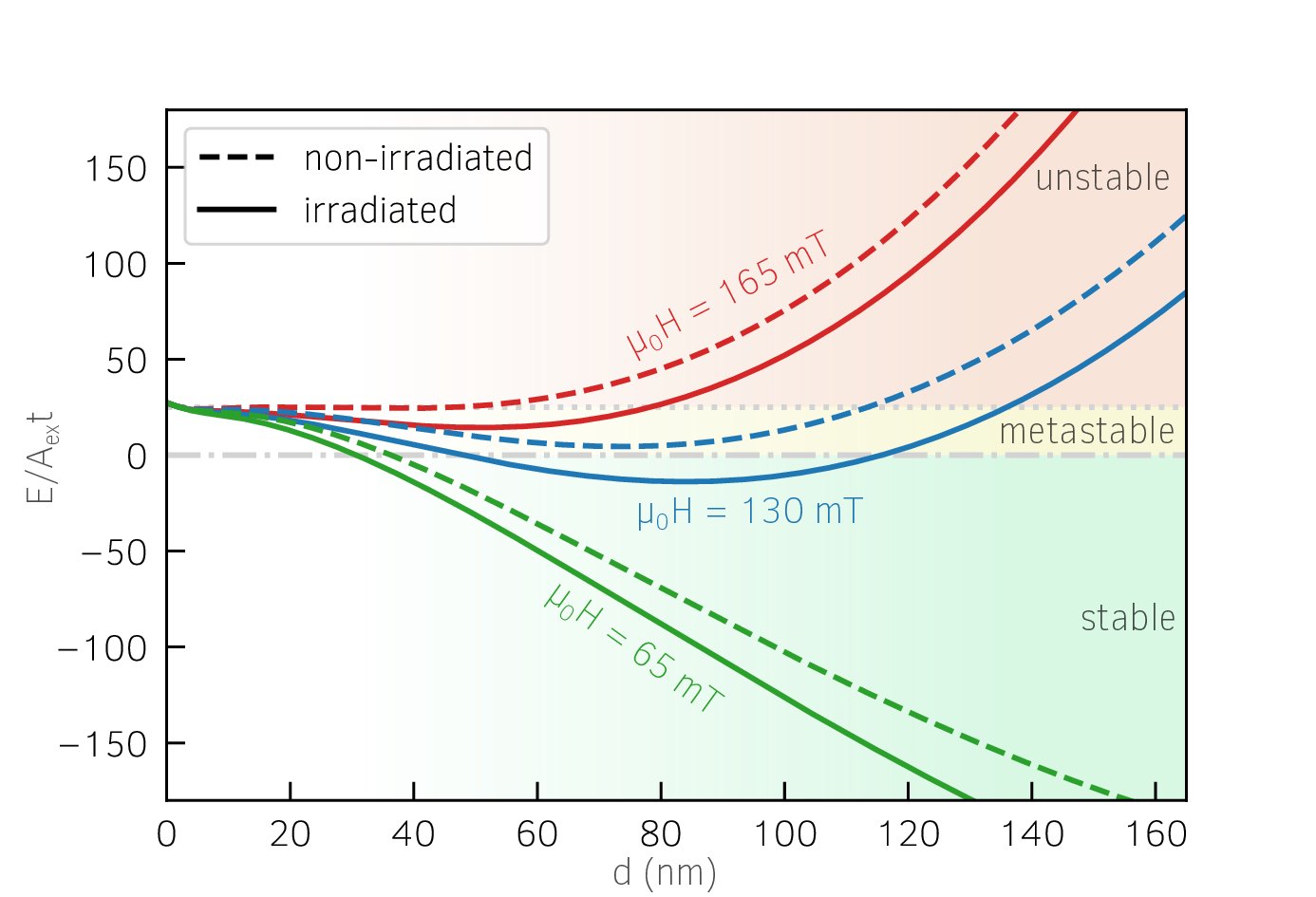}
	\caption{\textbf{Quasi-particle prediction of spin texture localization.} Energy $E$ of an isolated skyrmion as a function of its diameter, in units of $A_{\rm ex} t$ (exchange constant $A_{\rm ex}$ and film thickness $t$). Solid lines: irradiated areas. Dashed lines: non-irradiated film. Curves are shown for applied fields that correspond best to the experiment. Colored regions indicate stable (global minimum, $E<0$), metastable (local minimum, $E>0$), and unstable (no minimum) skyrmions. Skyrmions are expected to exist in material regions where a global minimum occurs. For \qty{65}{mT}, both curves have a global minimum outside the plotted range.
		\label{fig:5}}
\end{figure}

Figure~\ref{fig:5} shows the calculated energy–diameter relations for three representative magnetic fields, comparing irradiated and non-irradiated regions. Model parameters were obtained from room-temperature magnetic characterization (see Methods). As a general trend, skyrmion energy decreases with ion irradiation and increases with applied field. Consequently, at low field (65 mT), skyrmions are stable in both regions; at intermediate field (130 mT), stability remains only in irradiated areas; and at high field (160 mT), all skyrmions become metastable or unstable---closely matching the three experimentally observed regimes. Quantitatively, the model requires slightly higher fields than the experiment, likely due to the omission of inter-skyrmion stray-field interactions, which also lead to an overestimation of skyrmion diameters. This deviation diminishes at larger fields as the inter-skyrmion spacing increases. Aside from these expected differences, the model captures the localization conditions with remarkable precision. Notably, it does so using only room-temperature parameters, indicating that localization occurs once the system has cooled nearly back to ambient temperature.

\section{Unified interpretation of nucleation and localization dynamics} 

Our combined experimental and simulation results establish a coherent picture of picosecond nucleation and localization of nanometer-scale magnetic textures following ultrafast laser excitation. When the pulse drives the system toward the paramagnetic regime, thermal spin fluctuations become strong enough to compete with exchange interactions, thereby lowering the topological energy barrier and enabling the spontaneous nucleation of atomic-scale domains such as skyrmions. In this fluctuation state, weaker contributions---such as Zeeman coupling, anisotropy variations---play no significant role, consistent with the reported field-independence of this regime~\cite{gerlingerRobustScenarioGeneration2022}. As the system cools and exchange interactions recover, fluctuations subside and the nascent textures expand to mesoscopic scale. This transition marks the crossover from atomistic physics to a micromagnetic regime in which subtle energy balances and nanometer-scale material variations determine whether individual textures collapse or persist. Two experimental observations support this interpretation: localization begins exactly when the fluctuation state terminates, and the resulting configuration is in quantitative agreement with micromagnetic predictions for the local energetic ground state.

The interpretation of localization presented here---rooted in spin-texture decay---extends our earlier quasi-particle framework for ultrafast skyrmion nucleation~\cite{liefferinkEffectiveTheoryUltrafast2025} to the subsequent localization dynamics. Its strength lies in its ability to account for all experimentally observed features in an intuitive, coarse-grained picture that quantitatively predicts the final states using only statically accessible micromagnetic parameters. Because our measurements are not unambiguously sensitive to lateral spin currents, we make no claim regarding their presence or relevance, rendering our interpretation fully complementary to earlier reports of magnon-mediated localization on few-picosecond timescales~\cite{iacoccaSpincurrentmediatedRapidMagnon2019}. In our description, motion is required only for the small corrective displacements during the rearrangement phase, consistent with a single-particle relaxation picture~\cite{buttnerDynamicsInertiaSkyrmionic2015}. By abstracting away nonlocal and correlated dynamics, the quasi-particle framework provides a transparent and predictive understanding of ultrafast switching. Quantitatively, it reveals extraordinarily steep lifetime gradients---from sub-nanosecond in pristine regions (from SAXS) to hours within irradiated dots (from STXM)---spanning more than twelve orders of magnitude across less than \qty{100}{nm}. The ability to engineer such extreme lifetime contrasts, and to harness them for deterministic picosecond-scale localization, may represent the most practically impactful outcome of this work.

Finally, we note that the localization dynamics observed here are fundamentally different from the behavior expected in conventional phase transitions at slower timescales. In both first- and second-order transitions, localization induced by spatial inhomogeneities is well established, but it typically occurs \emph{during} nucleation. A textbook first-order example is solidification, where nucleation at heterogeneities fixes grain positions at the very onset of the transition. Current- or field-driven magnetic switching—the conventional analog to the transition studied here—likewise follows a localization-during-nucleation pathway. Even in second-order transitions, such as when cooling through the Curie temperature, localization is expected to emerge during nucleation: the inhomogeneous Kibble–Zurek mechanism predicts that small variations in the critical temperature pin the nascent domain pattern to regions of highest critical temperature as the system crosses the phase boundary sufficiently slowly~\cite{delcampoInhomogeneousKibbleZurek2011}. The clear temporal separation between nucleation and localization at ultrafast timescales revealed in this work is therefore remarkable and motivates further systematic investigation from a phase-transition perspective.

\section{Conclusions}

In conclusion, we presented a combined experimental and theoretical study that advances the understanding of ultrafast phase transitions into real space. We demonstrated that real-space dynamics become experimentally accessible to time-resolved Fourier-space probes when the energy landscape is periodically modulated at the nanometer scale. Using this approach, we tracked the nucleation and localization of nanometer-scale magnetic textures following ultrafast laser excitation and showed that both processes complete on a sub-nanosecond timescale. Remarkably, localization sets in only after nucleation has finished, distinguishing ultrafast phase transitions from their conventional counterparts. Mechanistically, nucleation is governed by atomic-scale fluctuations near the Curie temperature, dominated by exchange and chiral interactions, whereas the subsequent localization proceeds as a micromagnetic process near room temperature, shaped by the subtle balance and spatial modulation of competing micromagnetic energies. Strikingly, these competing interactions generate extreme lateral variations in spin-texture lifetimes, spanning from sub-nanoseconds to beyond hours over distances below \qty{100}{nm}. By demonstrating nanometer-scale spatial control on ultrafast timescales and establishing a multiscale framework to describe the underlying dynamics, this work provides a broadly applicable foundation for exploring nonequilibrium phenomena in condensed matter and for advancing ultrafast control in nanoscale technologies.

\FloatBarrier
\newpage

%%%%%%%%%%%%%%%%%%%%%%%%%%%%%%%%%%%%%%%%%%%%%%%%%%%%%%%%%%%%%%%%%%%%%%%%%%%%%%%%%%%%%%%%%%%%%%%%%%%%%%%%%%%%%%%%%%%%
\section{Methods}\label{sec:methods}
%%%%%%%%%%%%%%%%%%%%%%%%%%%%%%%%%%%%%%%%%%%%%%%%%%%%%%%%%%%%%%%%%%%%%%%%%%%%%%%%%%%%%%%%%%%%%%%%%%%%%%%%%%%%%%%%%%%%

\subsection{Sample growth}

This study uses the same sample set as our previous work on laser-induced topological switching~\cite{buttnerObservationFluctuationmediatedPicosecond2021}. The multilayer stack was Ta(\qty{3}{nm})/[Co(\qty{0.6}{nm})/Pt(\qty{0.8}{nm})]$_{\times15}$/Ta(\qty{2}{nm}), deposited by confocal d.c. magnetron sputtering (r.f. sputtering for Ta) at a base pressure of \qty{6.5e-9}{mbar} and an Ar working pressure of \qty{2.7e-3}{mbar}. Films were grown on \qty{150}{nm}-thick \ce{Si3N4} membrane chips with $\qty{200}{\micro\meter}\times\qty{200}{\micro\meter}$ x-ray–transparent windows. Layer thicknesses were calibrated using a quartz crystal monitor. Unless noted otherwise, all measurements were performed on the same sample.

\subsection{Ion irradiation}

\begin{figure}[p!]
	\centering
	\includegraphics[]{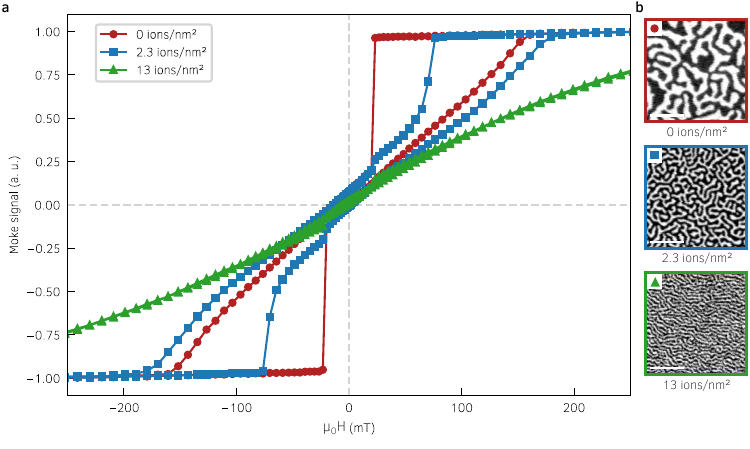}
	\caption[Characterization of the irradiated sample.]{\textbf{Characterization of the irradiated sample.} \textbf{a} Out-of-plane hysteresis loops recorded with Kerr-microscopy within $15\times\qty{15}{\micro\meter\squared}$-large regions irradiated at the specified Ga-irradiation dose. The values correspond to the pristine film (red), the dose used in the time-resolved experiment (\qty{2.3}{ions\per\nano\meter\squared}, blue) and a significantly higher dose (\qty{13}{ions\per\nano\meter\squared}, green).  \textbf{b} Domain state at \qty{0}{mT} applied field for each dose. The XMCD contrast images, showing the out-of-plane magnetization, were recorded with STXM. Scalebar \qty{1}{\micro\meter}.}
	\label{fig:xfel-material}
\end{figure}

Irradiated sites with a lattice period of \qty{400}{nm} and a dot diameter of \qty{100}{nm} were patterned using a focused ion beam microscope with Ga ions (ThermoFisher HELIOS 600). We used the maximum acceleration voltage of \qty{30}{keV}, which provides the most homogeneous penetration through the multilayer stack. Owing to the strong interlayer exchange coupling in the Co/Pt system, the film is expected to remain effectively homogeneous along the film normal even if ion-induced intermixing is stronger in the upper layers. A dose of \qty{2.3}{ions\per\nano\meter\squared} was selected because it significantly modified the nucleation and saturation fields compared to the pristine material while preserving the qualitative shape of the sheared hysteresis loop (Extended Data Fig.~\ref{fig:xfel-material}). The dot diameter was chosen to match the stripe-domain period of the irradiated material (\qty{90}{nm}, Extended Data Fig.~\subref{fig:xfel-material}{b}), ensuring that a single magnetic texture can form in each dot.

\subsection{X-ray microscopy}

Real-space x-ray imaging was performed with STXM at the Maxymus end station~\cite{weigandTimeMaxyneShotNoiseLimited2022} at the BESSY II electron storage ring operated by the Helmholtz-Zentrum Berlin für Materialien und Energie. We utilized the XMCD contrast at the $L_3$-edge of cobalt to resolve the out-of-plane magnetization structure of the magnetic domain states. Excitation pulses were provided by a fiber-incoupled Yb:KGW-laser with variable pulse energy and a pulse duration of $\qty{1}{ps}$.

\subsection{Determination of the topology of laser-nucleated magnetic textures}

We performed Lorentz TEM imaging at the Göttingen Ultrafast Transmission Electron Microscope to resolve the topology of the circular magnetic textures formed on the irradiated dots. The sample was excited in situ using an in-coupled laser integrated into the microscope. For each applied magnetic field, images of the initial and final states were acquired, and difference images were computed to suppress any non-magnetic background. Topological classification of the nucleated textures was carried out manually by inspecting the Lorentz contrast of each feature and comparing it to the characteristic signatures expected for Bloch-type skyrmions~\cite{heiglDipolarstabilizedFirstSecondorder2021, hassanDipolarSkyrmionsAntiskyrmions2024}.

\subsection{Time-resolved SAXS}

\begin{figure}[p!]
	\centering
	\includegraphics[]{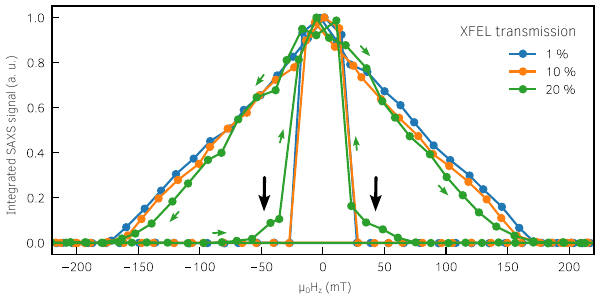}
	\caption[SAXS hysteresis curves.]{\textbf{SAXS hysteresis curves for various intensities of the XFEL pulse.} The curves show the frame-integrated total scattering intensity vs. the applied magnetic field for the specified photon transmission through a gas attenuator upstream of the sample. Small green arrows indicate the direction of the field sweep. At high fluence, the intense XFEL pulses start to alter the nucleation fields, as indicated by black arrows. \label{fig:xfel-FELintensity}}
\end{figure}

Scattering patterns were recorded using a Teledyne PI-MTE3 in-vacuum CCD detector equipped with a wire-mounted beamstop to block the direct (zeroth-order) transmission. An in-situ electromagnet provided the out-of-plane magnetic field. The x-ray probe pulse intensity was attenuated such that it did not perturb the magnetic state; the perturbation threshold was determined by recording SAXS hysteresis loops at varying FEL intensities, where intensities above threshold produced a shift of the spontaneous nucleation field toward higher values (Extended Data Fig.~\ref{fig:xfel-FELintensity}).

To reduce spatial jitter of the FEL beam, an upstream pinhole ($\diameter = \qty{35}{\micro\meter}$) was inserted to transmit only a small portion of each pulse. This procedure ensured the illumination of the sample with a stable, Airy-disk wavefront.

Optical excitation was provided by a femtosecond laser (OL) with wavelength $\lambda = \qty{800}{nm}$, pulse duration $w \approx \qty{50}{fs}$, spot size $\diameter_{\text{FWHM}} = \qty{260}{\micro\meter}$, and fluence $F_{\text{FWHM}} = \qty{17.6}{mJ/cm^2}$, synchronized to the FEL with an adjustable delay. Significant shot-to-shot spatial jitter and profile variations of the OL led to fluence fluctuations of up to \qty{100}{\percent} at the probed region. To account for this, the effective pump fluence was determined for each shot by recording a virtual image of the optical spot using an external camera during each pump–probe event.

Each pump–probe cycle (Extended Data Fig.~\subref{fig:xfel-procedure}{b}) consisted of:
\begin{enumerate}
\item Applying an out-of-plane field of $B_{\text{reset}}=\qty{300}{mT}$ to reset the sample to a uniformly magnetized state.
\item Reducing the field to $B_{\text{meas}}$, the value at which the pump–probe event was taken.
\item Exciting the sample with the OL pulse.
\item Probing the magnetic state with a single FEL pulse after a delay $\tau$ and recording the associated SAXS pattern.
\item During detector readout and sensor preparation, repeating steps 1 and 2 to prepare for the next event.
\end{enumerate}

The FEL facility was operated in a pulse-on-demand mode, where an X-ray pulse can be delivered every \qty{0.1}{s}, if requested from the experimentalist. The experimental timing was configured such that each complete pump–probe sequence required \qty{0.3}{s}, corresponding to three FEL periods, ensuring reliable trigger synchronization and sufficient time for magnetic-field cycling and detector readout. Each delay point was sampled with 400--800 single-shot pump–probe events, with all frames stored individually for single-shot-level analysis.

Quasi-static data of the final states (Fig.~\ref{fig:3}) were recorded with 48 pulses per cycle, leading to 48 times stronger overall signal per camera frame.

\subsection{Data treatment}

Some regions of the scattering patterns were obscured by camera defects, the beamstop, and several bright square spots arising from direct x-ray transmission through neighboring membrane windows. The following post-processing steps were applied to the raw data to correct for these effects and extract the presented scattering data.

\textbf{Thresholding:} Pixels without photon hits contribute only electronic read noise and dark current to shot-averaged images. To suppress this bias in the low-count regime, each frame was photon-discriminated by a per-quadrant threshold derived from the noise histogram of unexposed pixels. The threshold was set to the mean of this histogram plus 0.8 times its standard deviation. Sub-threshold pixels were set to zero before averaging.

\textbf{Azimuthal averaging and separation of signal contributions:} 
Each frame contained a combination of Bragg scattering $I_{\text{Bragg}}(\mathbf{q},t)$, isotropic scattering $I_{\text{iso}}(\mathbf{q},t)$ and artifacts.
Bragg peak positions were determined by manually selecting four different Bragg positions, extracting their scattering order and base vectors, and using these to calculate the remaining peak positions. We omitted an Ewald projection of the data as with a maximum scattering angle of \qty{6.2}{\degree}, the corresponding corrections are below our pixel resolution and do not impact the analysis.
All Bragg peaks, as well as the areas with parasitic stray light and the beamstop-covered parts of the CCD, were masked. The remaining image contained only isotropic scattering.
Exploiting the isotropic symmetry, the masked areas were filled in with the azimuthal average of the scattering intensity for each $q$ to get an approximation of $I_{\text{iso}}(\mathbf{q},t)$ (top-right sections in Fig.~\ref{fig:3}).
$I_{\text{Bragg}}(\mathbf{q},t)$ was obtained by subtracting the calculated purely isotropic scattering from the original SAXS pattern (bottom-right sections in Fig.~\ref{fig:3}). 

\textbf{Normalization and filtering by optical excitation strength:}
Fluctuations in X-ray intensity were corrected by normalizing the scattering images to the direct X-ray transmission through neighboring membranes (see Extended Data Fig.~\subref{fig:xfel-procedure}{c}). To ensure that the optical laser deposited sufficient energy to drive the system above the nucleation threshold~\cite{gerlingerApplicationConceptsUltrafast2021}, frames were further filtered based on the measured laser excitation strength. By analyzing the monitored laser positions for pulses that produced strong SAXS signals, we identified a region on the laser monitor corresponding to the sample location. The intensity in this region was thus taken as proportional to the actual laser fluence on the magnetic sample.

\begin{figure}[p!]
	\centering
	\includegraphics[]{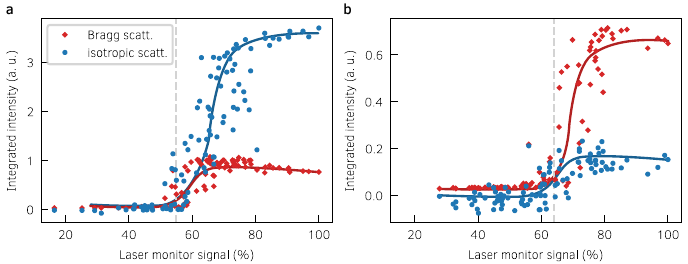}
	\caption{\textbf{Dependence of final-state scattering intensity on optical excitation strength.} The integrated isotropic and Bragg contributions of the 100 brightest individual final state scattering frames are shown. \textbf{a} SAXS contributions at a low applied field of \qty{65}{mT}. \textbf{b} Scattering contributions at an intermediate applied field of \qty{115}{mT}. Dashed lines indicate the minimum required laser intensity for nucleation. Solid lines are guides for the eye.\label{fig:xfel-laser}}
\end{figure}

Extended Data Fig.~\ref{fig:xfel-laser} shows the final scattering intensity for 100 individual CCD frames as a function of the estimated laser intensity. Above a field-dependent threshold, scattering is reliably observed. Any frames with a laser monitor signal below this threshold were discarded.

\subsection{Extraction of skyrmion size}

\begin{figure}[p!]
	\centering
	\includegraphics[]{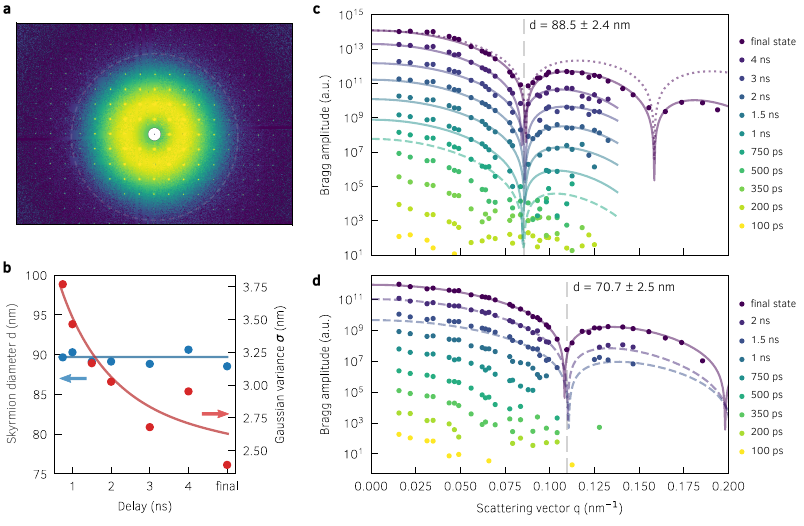}
	\vspace{-0.2cm}
	\caption{\textbf{Spin texture size estimation from order-resolved Bragg scattering.} \textbf{a} Final-state SAXS pattern at low field showing a radial minimum in the Bragg intensity (dashed line). \textbf{b} Fits of the $q$-dependent Bragg intensity yield the skyrmion diameter $d$ and a spatial coherence estimate $\sigma$. Solid lines are guides to the eye. \textbf{c},\textbf{d} Average Bragg intensity for each recorded delay at low (\textbf{c}) and medium (\textbf{d}) field, plotted logarithmically, with offsets for clarity. Lines show fits using an Airy disk multiplied by a Gaussian envelope. Solid lines indicate cases where the Airy minimum is clearly observed, dashed lines where the minimum is not visually verifiable due to low signal-to-noise. The dotted line shows the expected Bragg amplitude without the Gaussian envelope, assuming perfectly aligned magnetic textures. The first Airy minimum is indicated with gray dashed lines and the corresponding spin texture diameter is $d$ stated. Gray dashed lines mark the first Airy minimum, with the corresponding spin texture diameter $d$ indicated. The uncertainty in $d$ reflects the fit error (standard deviation of the mean), not the spread of individual diameters.}
	\label{fig:xfel-airy}
\end{figure}

The intensity of individual Bragg peaks is influenced by the average shape, size, and arrangement of all skyrmions contributing to the SAXS pattern. In particular, the finite real-space extent of the skyrmions acts as a convolution, producing a $\mathbf{q}$-dependent modulation of the Bragg intensity in Fourier space. Extended Data Figures~\subref{fig:xfel-airy}{c,d} show the average intensity of all Bragg peaks with the same $q = |\mathbf{q}|$ for different time delays and for low and intermediate applied fields.

At long delays and in the final states, the scattering patterns exhibit radial intensity minima characteristic of Airy patterns, which arise from diffraction by circular features. The position of the first minimum directly reflects the feature diameter.

To extract the skyrmion diameter, we fitted the Airy intensity
\begin{align}
I_{\text{Airy}} = I_0 \left(\frac{2 J_1(\gamma)}{\gamma}\right)^2,
\end{align}
with $\gamma = \frac{\pi d}{\lambda} \sin\theta$ and $\theta = 2 \arcsin\left(\frac{q \lambda}{4 \pi}\right)$ \cite{pedrottiIntroductionOptics2007}, to each dataset. Here, $J_1$ is the first-order Bessel function, and the fit parameters were the intensity scale $I_0$ and the skyrmion diameter $d$. While pure Airy patterns accurately describe the intensity profiles at low $q$, they fail to capture the decay at larger $q$ (see final state in Extended Data Fig.~\subref{fig:xfel-airy}{a}). This deviation arises because perfect Airy functions assume circular features precisely aligned on a grid.

To account for imperfect positioning and deviations from ideal circularity due to finite domain wall widths, we multiplied the fit function by a Gaussian envelope, $e^{-\sigma \gamma^2}$. Importantly, this Gaussian modulation does not shift the Airy minimum, allowing robust extraction of the skyrmion diameter $d$. Consequently, reliable diameters are reported only for scattering patterns with clearly observable Airy minima, where the Gaussian width $\sigma$ can also be accurately determined.

\subsection{Atomistic spin dynamics simulations}

Atomistic simulations were performed to support the experimental results. Extending on previous studies \cite{buttnerObservationFluctuationmediatedPicosecond2021}, we consider a spin model described by the Hamiltonian
\begin{equation}
    H=- \sum_{ij} J_{ij} \left( \mathbf{m}_i \cdot \mathbf{m}_j \right) - \sum_{ij} \mathbf{D}_{ij} \cdot \left( \mathbf{m}_i \times \mathbf{m}_j \right) - \sum_{i} K_i m_{i,z} ^2 - \sum_{i} M_s \left( m_{i,z} B_z \right).
\label{eq:Hamiltonian}
\end{equation}
Here, $\mathbf{m}_i = \mathbf{M}_i/M_s$ is the normalized magnetic moment at lattice site $i$. $m_{k,z}$ is the out-of-plane component and $M_s=4.66\mu_B$. The  exchange interaction $J_{ij}$ is non-zero only for nearest neighbors, and has a magnitude of $J=\qty{1.5}{meV}$. $\mathbf{D}_{ij}$ is the vector describing the DMI between site $i$ and $j$, with a vector orientation that favors N\'eel type skyrmions and a magnitude of $D=\qty{0.6}{meV}$. The out-of-plane magnetic field strength is $B_z=\qty{0.5}{T}$. Simulations were performed on a square lattice with side lengths of $L=500$ atomic spacings, with periodic boundary conditions. To model the irradiated dots, we employed a site-dependent anisotropy
\begin{equation}
    K_i = \begin{cases}
        K \qquad \quad \text{ if} \qquad i \notin \text{dot} \\
        K/10\qquad \text{if} \qquad i \in \text{dot}

    \end{cases}
\end{equation}
with $K=\qty{0.93}{meV}$. We used 25 dots, distributed evenly as a $5\times5$ square pattern. Each dot has $16\times 16$ lattice points, approximately matching the size of the final-state skyrmions. Anisotropies were chosen such that outside the dots, skyrmions decay after the heat pulse, while inside the dots, skyrmions mostly remain stable for the duration of the simulation \cite{kernDeterministicGenerationGuided2022}. 
We simulated the dynamics by solving the stochastic Landau-Lifshitz-Gilbert equation
\begin{equation}
    \frac{ \mathrm{d} \mathbf{m}_i}{\mathrm{d}t}= - \gamma \mathbf{m}_i \times \mathbf{B}_i + \frac{\alpha}{M_s} \mathbf{m}_i \times \frac{ \mathrm{d} \mathbf{m}_i}{ \mathrm{d}t }
    \label{eq:LLG},
\end{equation}
where $\gamma$ is the electron gyromagnetic ratio and $\alpha=0.3$ is the Gilbert damping. $\mathbf{B}_i$ is the effective magnetic field at site $i$, and is given by
\begin{equation}
    \mathbf{B}_i = -\frac{\partial H}{\partial \mathbf{m}_i} + \boldsymbol{\zeta}(t) 
\end{equation}
where $\boldsymbol{\zeta}(t)$ is a stochastic field described by uncorrelated Gaussian white noise with zero mean and a variance given by the fluctuation-dissipation theorem,
\begin{equation}
    \left \langle \zeta_i^{\mu} \left( t \right) \zeta_j^{\nu} \left( t' \right) \right \rangle = 2 \alpha \frac{k_B T}{\gamma M_s} \delta_{i,j} \delta_{\mu,\nu} \delta \left(t-t' \right),
\end{equation}
where $T$ is the temperature of the heat bath and $\mu,\nu$ denote Cartesian components. We model the effect of the laser pulse by simulating the system in response to a time-dependent temperature which rises and decays exponentially on top of a base temperature $T_0=\qty{4}{K}$:
\begin{equation}
    T(t) = T_0 + T_{\text{pulse}}(1-e^{-t/t_1})e^{-t/t_2},
\end{equation}
where $T_{\text{pulse}}= \qty{23}{K}$, $t_1=\qty{1}{ps}$, and $t_2=\qty{30}{ps}$.
The system of equations in Eq. \eqref{eq:LLG} is solved by using the UppASD code (v5.1.1) \cite{skubicMethodAtomisticSpin2008, UppASDKTHProject}, with the semi-implicit time-integration method described in \cite{mentinkStableFastSemiimplicit2010}. To achieve numerical convergence, we used a step size of $\Delta t = \qty{50}{fs}$. The simulation was initiated in the ferromagnetic ground state and run for a total of 2000 time steps, where every 20th step was saved.

To gauge how many skyrmions we have in the system at any given time, we calculate the topological charge. This is done by splitting the lattice into nearest neighbor triangles $i$ \cite{bergDefinitionStatisticalDistributions1981}, and calculating the local topological charge $q_i$ of that triangle, given by
\begin{equation}
    q_i = 2 \tan^{-1}\left( \frac{\mathbf{m}_{i,1}\cdot(\mathbf{m}_{i,2}\times\mathbf{m}_{i,3})}{1 + \mathbf{m}_{i,1}\cdot\mathbf{m}_{i,2} + \mathbf{m}_{i,1}\cdot \mathbf{m}_{i,3}+\mathbf{m}_{i,2}\cdot\mathbf{m}_{i,3}} \right).
\end{equation}
The total topological charge is then the sum $Q=\sum_i q_i$ of the contributions from every triangle.

\subsection{Micromagnetic parameter estimation}

To achieve an analytical model close to the investigated Co/Pt-multilayer, we performed measurements to acquire a set of micromagnetic parameters. 
Besides the spatial structure of the film, these include the exchange stiffness~$A_{\text{ex}}$, the saturation magnetization~$M_{\text{s}}$, the DMI strength $D$ and the magnetic perpendicular anisotropy constant~$K_{\text{u}}$. 
It is difficult to extract all parameters from the same sample, especially in case of local ion irradiation (examples from literature require, e.\,g., the creation of microscopic Hall crosses \cite{dejongLocalControlMagnetic2022}).
We therefore obtained the parameters by measuring different properties on two multilayer samples (the actual membrane sample and a nominally identical reference multilayer on a silicon substrate that was fabricated in the same sputtering system around the same time), in combination with a set of crucial assumptions:
\begin{enumerate}
	\item The saturation magnetization, the exchange stiffness and the DMI do not vary between these two samples.
	\item The exchange stiffness is expected to be in the range of \qtyrange{10}{25}{pJ/m} \cite{legrandSpatialExtentDzyaloshinskiiMoriya2022}. 
	\item Flux closure domain walls are expected in the material, which we model by allowing a finite DMI strength \cite{dovzhenkoMagnetostaticTwistsRoomtemperature2018}.
	\item Ion irradiation affects only the local perpendicular anisotropy.
\end{enumerate}

\begin{figure}[p!]
	\centering
	\includegraphics[]{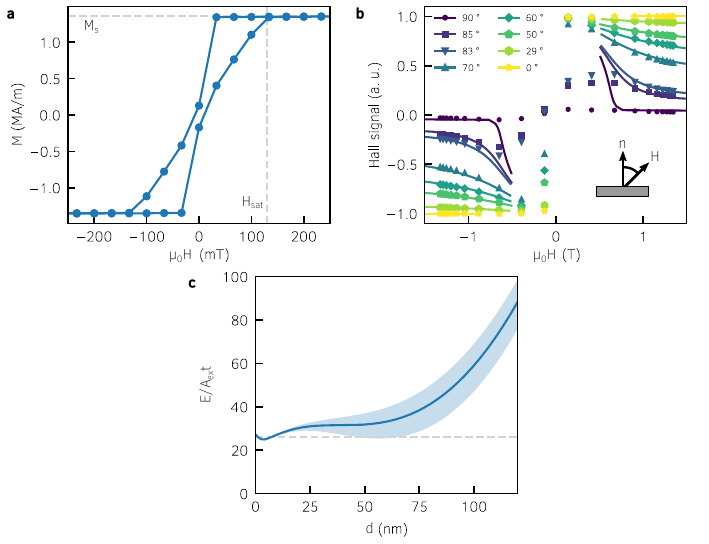}
	\caption{\textbf{Magnetic characterization of the material.} \textbf{a} Out-of-plane hysteresis loop (SQUID) of a reference sample. Magnetization is normalized to the total Co thickness. \textbf{b} Hall signal proportional to the out-of-plane magnetization vs. field of the same reference sample for various applied field angles. Solid lines show the fit result of a single-domain Stoner Wohlfahrt model, allowing for the extraction of the uniaxial anisotropy $K_\text{u}$. \textbf{c} Skyrmion energy curve at the saturation field $H_{\text{sat}}$, at which the local minimum vanishes. The shaded area depicts the range caused by a \qty{10}{\percent} variance in the material parameters, exemplary shown for the exchange stiffness. \label{fig:xfel-parameterDetermination}}
\end{figure}

We measured the saturation magnetization of the reference sample via SQUID magnetometry in a Quantum Design MPMS3-setup. 
The out-of-plane hysteresis loop was recorded and the linear diamagnetic background subtracted. We determined the saturation magnetization $M_{\text{s}}=\qty{1.35}{MA/m}$ as well as the saturation field $H_{\text{sat}}=\qty{130}{mT}$ (Extended Data Fig.~\subref{fig:xfel-parameterDetermination}{a}).

We used Stoner-Wohlfarth analysis to determine the magnetic anisotropy $K_{\text{u,pristine}}$ of the same reference film.
We recorded hysteresis curves of the out-of-plane magnetization for various different angles between the sample and the applied field with a zero-offset-corrected Hall-transport setup (see Extended Data Fig.~\subref{fig:xfel-parameterDetermination}{b}).
In the high field regime ($\mu_0 H>\qty{1}{T}$) no domains exist and the magnetization rotates coherently. In consequence, the Stoner-Wohlfarth model description solely requires the saturation magnetization to determine the effective anisotropy of the film.
Fitting the dataset according to the procedure described in \cite{lavrijsenAnotherSpinWall2011}, we acquired a ratio of $K_{\text{eff}}/M_{\text{s}}=\qty{1.6}{J/Am^2}$.
With the known $M_{\text{s}}$, this result corresponds to $K_{\text{u}}=\qty{1.57}{MJ/m^3}$.

With the saturation magnetization and anisotropy known, we estimated the strength of exchange and DMI interaction from the saturation field.
We used the analytical skyrmion model \cite{buttnerTheoryIsolatedMagnetic2018} to find a set of parameters which predict a skyrmion collapse for the measured saturation field~$H_{\text{sat}}$. 
In the skyrmion model, this condition is reflected by the disappearance of a local energy minimum at $H_{\text{sat}}$.
Extended Data Figure~\subref{fig:xfel-parameterDetermination}{c} depicts this condition for the reference film. 
We find $D=\qty{1.5}{mJ/\meter\squared}$ and $A_{\text{ex}}=\qty{12}{pJ/m}$, respectively.
These values are chosen to represent a realistic multilayer stack, due to their strong covariance in the model (a larger $D$ also results in a larger $A_{\text{ex}}$).

Finally, in a similar manner the anisotropies of the irradiated and non-irradiated sample were determined.
While the reference film has a saturation field of $H_{\text{sat}}=\qty{130}{mT}$, the non-irradiated multilayer exhibits saturation at $H_{\text{sat}}=\qty{175}{mT}$ and the irradiated patches at an even higher $H_{\text{sat}}=\qty{200}{mT}$ (compare MOKE curves in Fig.~\subref{fig:xfel-material}{a}).
By decreasing the perpendicular anisotropy, we can also shift the saturation field to higher values in our model: The domain wall surrounding the skyrmion is less energy intensive and thus a stronger external field is required to destabilize the skyrmion.
We determined anisotropy values of $K_{\text{u}} = \qty{1.35}{MJ/\meter\cubed}$ for the non-irradiated film and $K_{\text{u}} = \qty{1.25}{MJ/\meter\cubed}$ for the ion-irradiated areas.

\subsection{Skyrmion stability modeling}

The experimentally derived micromagnetic parameters were fed into an analytic model of a single, isolated skyrmion \cite{buttnerTheoryIsolatedMagnetic2018}, using the Wolfram Mathematica-based code available on Github~\cite{WolframCode}. Total energy curves versus skyrmion diameter were calculated for the given sets of micromagnetic parameters and applied magnetic out-of-plane fields. For each curve, the existence and position of a local energy minimum at finite skyrmion diameter were determined.

\section{Declarations}

\subsection{Acknowledgements}
We acknowledge European XFEL in Schenefeld, Germany, for provision of X-ray free-electron laser beamtime at SCS under proposal number 3463 and would like to thank the staff for their assistance. 
We acknowledge Tobias Freyermuth and Bernard Baranasic for developing the software to synchronize the electromagnet, the optical laser and the FEL.
We thank the Helmholtz-Zentrum Berlin für Materialien und Energie for the allocation of synchrotron radiation beamtime.
We acknowledge the use of the Physical properties laboratory, which is part of the CoreLab “Quantum Materials” operated by the Helmholtz-Zentrum Berlin.
Samples were manufactured at the TU Berlin Nano-Werkbank, which was supported by EFRE under contract no. 20072013 2/22.
We acknowledge funding by the Helmholtz Young Investigator Group Program through Project No. VH-NG-1520 and from the Deutsche Forschungsgemeinschaft (DFG, German Research Foundation) through Project No. 49254781 (TRR 360) subproject C02, Project No. 462676630  (BiSky), and Project No. 505818345 (Topo3D). B.S. acknowledges support by the European Research Council under the European Union’s Horizon 2020 research and innovation programme, within the Hidden, Entangled and Resonating Order (HERO) project with grant agreement 810451.

\subsection{Author contributions}

FB, DM and BP conceived the study.
SP, WDE, MS, RB, and DM fabricated the samples.
DM, StW and RB  characterized the samples with support from KP.
DM, MS, GM, TH, RB, CK, SW, MP, KPJ, VD, REC, LM, MT, AS, SZ, BVS, JS, KB, BP, and FB performed the experiments at XFEL.
DM, TS, RB, StW, and SW performed the STXM experiments at BESSY II with support from IW.
MM performed the LTEM experiments.
DM analysed the X-ray experiments with support from RB and FB.
TH performed the atomistic modeling with support from JHM, DM, and FB.
DM prepared the figures.
DM, FB and JHM wrote the manuscript with input from all authors.
Supervision was by CR, SE, JHM, BP, and FB.

\subsection{Competing interests}
The authors declare no competing interests.

\FloatBarrier

\bibliography{mybetterbib.bib,manualbib.bib}% common bib file

\makeatletter
\@fpsep\textheight
\makeatother

\FloatBarrier
\newpage
\section{Figures}
\newpage
\FloatBarrier

\FloatBarrier
\newpage

\FloatBarrier
\newpage

\FloatBarrier
\newpage

\FloatBarrier
\newpage

\FloatBarrier
\newpage

\FloatBarrier
\section*{Extended data figures}
\FloatBarrier

\renewcommand{\figurename}{Extended Data Fig.}

\BeginExtendedfigures

\newpage
\FloatBarrier

\FloatBarrier
\newpage

\FloatBarrier
\newpage

\FloatBarrier
\newpage

\FloatBarrier
\newpage

\FloatBarrier
\newpage

\FloatBarrier
\newpage

\FloatBarrier
\newpage

\EndExtendedfigures

\end{document}